\documentclass[]{spie}  

\usepackage{amsmath,amsfonts,amssymb}
\usepackage{graphicx}
\usepackage[colorlinks=true, allcolors=blue]{hyperref}

\title{Status of the Keck Planet Imager and Characterizer Phase II Development}

\author[a]{Jacklyn Pezzato}

\author[a]{Nemanja Jovanovic}
\author[a]{Dimitri Mawet}

\author[b]{Garreth Ruane}
\author[a]{Jason Wang}
\author[b]{James K. Wallace}
\author[a]{Jennah K. Colborn}
\author[d]{Sylvain Cetre}
\author[c]{Charlotte Z. Bond}
\author[b]{Randall Bartos}
\author[a]{Benjamin Calvin}
\author[a]{Jacques-Robert Delorme}
\author[a]{Daniel Echeverri}
\author[e]{Rebecca Jensen-Clem}
\author[e]{Eden McEwen}
\author[d]{Scott Lilley}
\author[d]{Ed Wetherell}
\author[d]{Peter Wizinowich}

\affil[a]{Department of Astronomy, California Institute of Technology, Pasadena, CA 91125, USA}
\affil[b]{Jet Propulsion Laboratory, California Institute of Technology, Pasadena, CA 91109, USA}
\affil[c]{Institute for Astronomy, University of Hawaii, 640 N. Aohoku Place, Hilo, HI 96720, USA}
\affil[d]{W. M. Keck Observatory, 65 - 1120 Mamalahoa Hwy., Kamuela, HI 96743, USA.}
\affil[e]{University of California, Berkeley, 510 Campbell Hall, Astronomy
Department, Berkeley, CA 94720, USA}

\authorinfo{Send correspondence to J. Pezzato \\ E-mail: jpezzato@caltech.edu}

\begin{document} 
\maketitle

\begin{abstract}
The Keck Planet Imager and Characterizer comprises of a series of upgrades to the Keck II adaptive optics system and instrument suite to improve the direct imaging and high resolution spectroscopy capabilities of the facility instruments NIRC2 and NIRSPEC, respectively. 
Phase I of KPIC includes a NIR pyramid wavefront sensor and a Fiber Injection Unit (FIU) to feed NIRSPEC with a single mode fiber, which have already been installed and are currently undergoing commissioning. 
KPIC will enable High Dispersion Coronagraphy (HDC) of directly imaged exoplanets for the first time, 
providing potentially improved detection significance and spectral characterization capabilities compared to direct imaging.
In favorable cases, Doppler imaging, spin measurements, and molecule mapping are also possible.
This science goal drives the development of phase II of KPIC, which is scheduled to be deployed in early 2020. 
Phase II optimizes the system throughput and contrast using a variety of additional submodules, including a 952 element deformable mirror, phase induced amplitude apodization lenses, an atmospheric dispersion compensator, multiple coronagraphs, a Zernike wavefront sensor, and multiple science ports. 
A testbed is being built in the Exoplanet Technology Lab at Caltech to characterize and test the design of each of these submodules before KPIC phase II is deployed to Keck. 
This paper presents an overview of the design of phase II and report on results from laboratory testing.
\end{abstract}

\keywords{Exoplanets, high contrast imaging, high contrast high resolution spectroscopy, small inner working angle coronagraphy, vortex coronagraph, on-axis segmented telescopes, apodization}

\section{INTRODUCTION}
\label{sec:intro}  

High Dispersion Coronagraphy (HDC) is a technique that optimally combines direct, high-contrast imaging techniques and medium-to-high resolution (R$\sim$1,000-100,000) spectroscopy \cite{Mawet2017, Wang2017}.
With modern telescopes, HDC will allow for the detection and high-resolution spectroscopic characterization of exoplanets and substellar companions.
The Keck Planet Imager and Characterizer (KPIC) will be the first instrument to enable HDC of directly imaged exoplanets \cite{Mawet2018}.


KPIC is a series of upgrades to Keck AO that will improve the performance of two preexisting Keck II facility instruments, NIRC2 and NIRSPEC, by incorporating a near-infrared pyramid wavefront sensor (PyWFS) and a fiber injection unit (FIU) \cite{Delorme2018, Mawet2018, Bond2018}.
The PyWFS both improves the direct imaging capabilities of NIRC2 and is key to successfully implementing HDC with the FIU.
This FIU feeds light from the Keck adaptive optics bench to NIRSPEC, an infrared echelle spectrograph, through a single mode fiber (SMF).

KPIC will achieve starlight suppression with apodizing coronagraphs, speckle nulling, and the SMF itself.
The fiber acts as a spatial filter by coupleing light on the fiber's axis.
By placing this fiber at the location of the planet, additional starlight is suppressed.
Dispersing the light to medium-to-high resolutions also avoids the chromatic speckle noise issues that hinder high-contrast low spectral resolution integral field units, and leverages the differences between the radial velocity and spectral signals of the planet and those of the star
\cite{Konopacky2013, Snellen2014, Barman2015, Wang2017, Mawet2017, Bryan2018}.
The improvements to contrast offered by these spectral differences relax the wavefront control requirements of the system, which allows for observations across large bandpasses.
Dispersing the light at medium-to-high resolutions also allows for the information content of the spectra to be optimally exploited using prior knowledge about the spectra of molecules of interest and the template matching cross-correlation technique.
Cross-correlation of the spectra increases the likelihood of a statistically significant detection of molecules and provides information about line widths that is useful for spin measurements and Doppler imaging \cite{Konopacky2013}.

The installation of KPIC at W. M. Keck Observatory is divided into three phases.
This approach allows for the system's performance to be optimized at each stage and will streamline the implementation of each round of system upgrades.
In addition to allowing for the verification of important low-level system functionalities before further improvements are installed, this approach also allows for the instrument to begin producing science results while upgrade paths are being explored in parallel.

The first phase of KPIC aims to deliver the Pyramid WFS and the FIU for early on-sky testing. The hardware has already been installed on the summit and the PyWFS and FIU are currently undergoing commissioning. See Jovanovic et al. 2019 (these proceedings) for more information on the current status of KPIC phase I.

Phase II of KPIC is driven entirely by the goal of demonstrating HDC of directly imaged exoplanets on-sky for the first time.
To achieve this goal, phase II will exploit various submodules 
designed to maximize the throughput and raw contrast of the system, therefore reducing the exposure time necessary to perform HDC.
The submodules include a 1K continuous-surface Deformable Mirror (DM), Phase Induced Amplitude Apodization (PIAA) lenses, an Atmospheric Dispersion Compensator (ADC), multiple coronagraphs, a Zernike wavefront sensor, and multiple fiber ports.

This proceedings will provide an overview of the design and status of the KPIC phase II developments, with a particular focus on four key 
submodules that are geared toward increasing the instrument's efficiency.
Section \ref{testbed} details the status of the KPIC phase II testbed, which is being built in the Exoplanet Technology Laboratory at the California Institute of Technology to streamline the implementation of phase II at Keck. 
Sections \ref{dm}, \ref{coronagraphs}, \ref{adc}, and \ref{piaa} focus on the design and status of the DM, coronagraphs, ADC, and PIAA submodules, respectively. 
Section \ref{conclusions} reviews the overall current status of KPIC phase II and the advantages offered by each of these submodules.

\section{KPIC phase II testbed}
\label{testbed}

\begin{figure} [ht]
    \begin{center}
    \begin{tabular}{c} 
    \includegraphics[height=8cm]{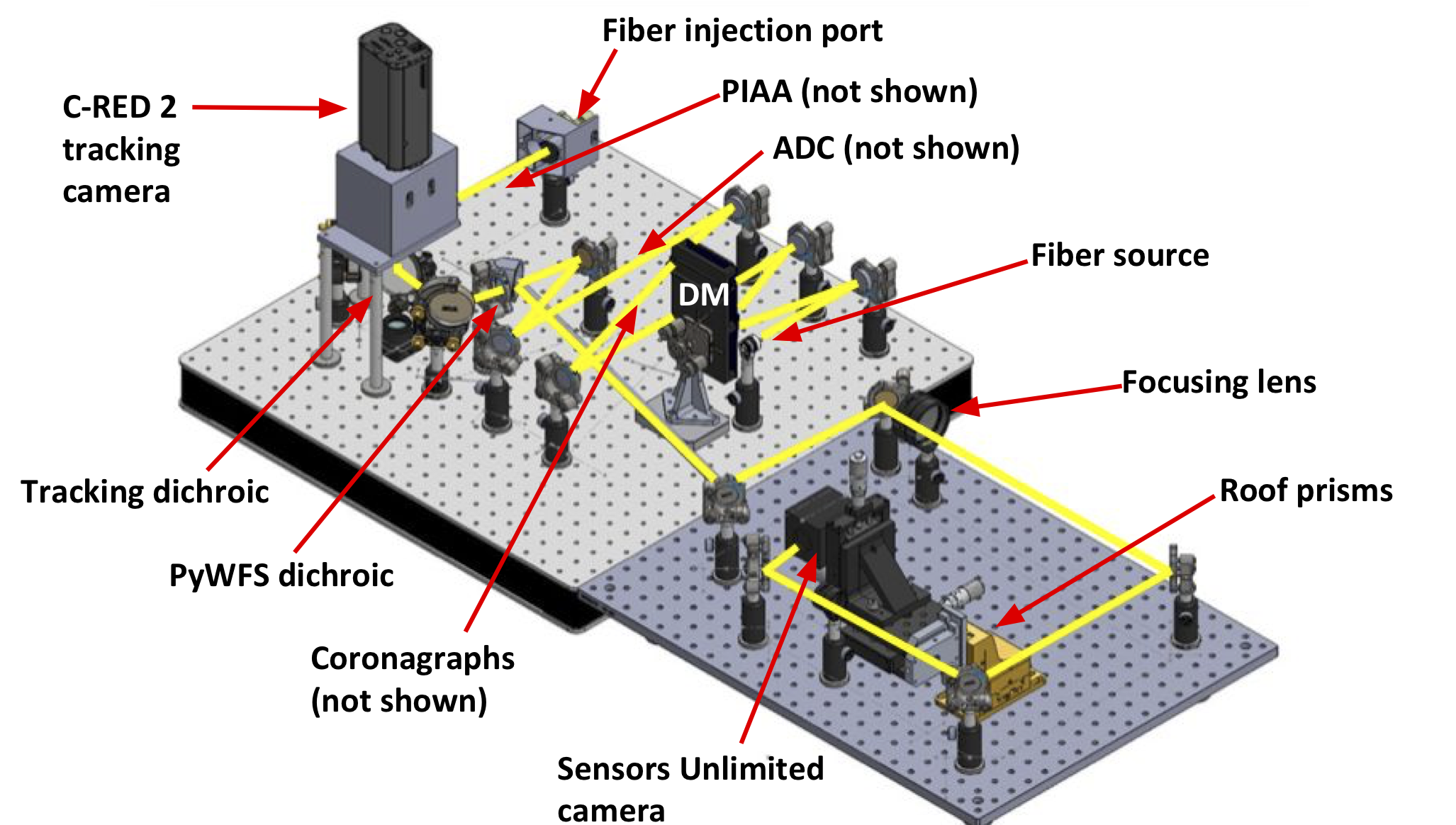}
    \end{tabular}
    \end{center}
    \caption[example] 
    { 
    \label{fig:testbed_cad} 
    CAD model of the KPIC phase II testbed, showing both the FIU plate (light gray, left) and the PyWFS plate (dark gray, right). 
    The optical path is shown in yellow.
    Light is injected into the testbed at the fiber source and passes the 1K DM in the first pupil plane.
    The coronagraphs submodule is in the second pupil and is immediately followed by the PyWFS dichroic.
    In transmission through the PyWFS dichroic is the tracking camera submodule, the PIAA submodule, and the fiber injection port.
    In the PyWFS arm, the beam passes through a long focal length focusing lens, the roof prisms, and finally through a collimating lens onto the Sensors Unlimited Camera. 
    }
\end{figure}

\begin{figure} [ht]
    \begin{center}
    \begin{tabular}{cc} 
    \includegraphics[height=5.6cm]{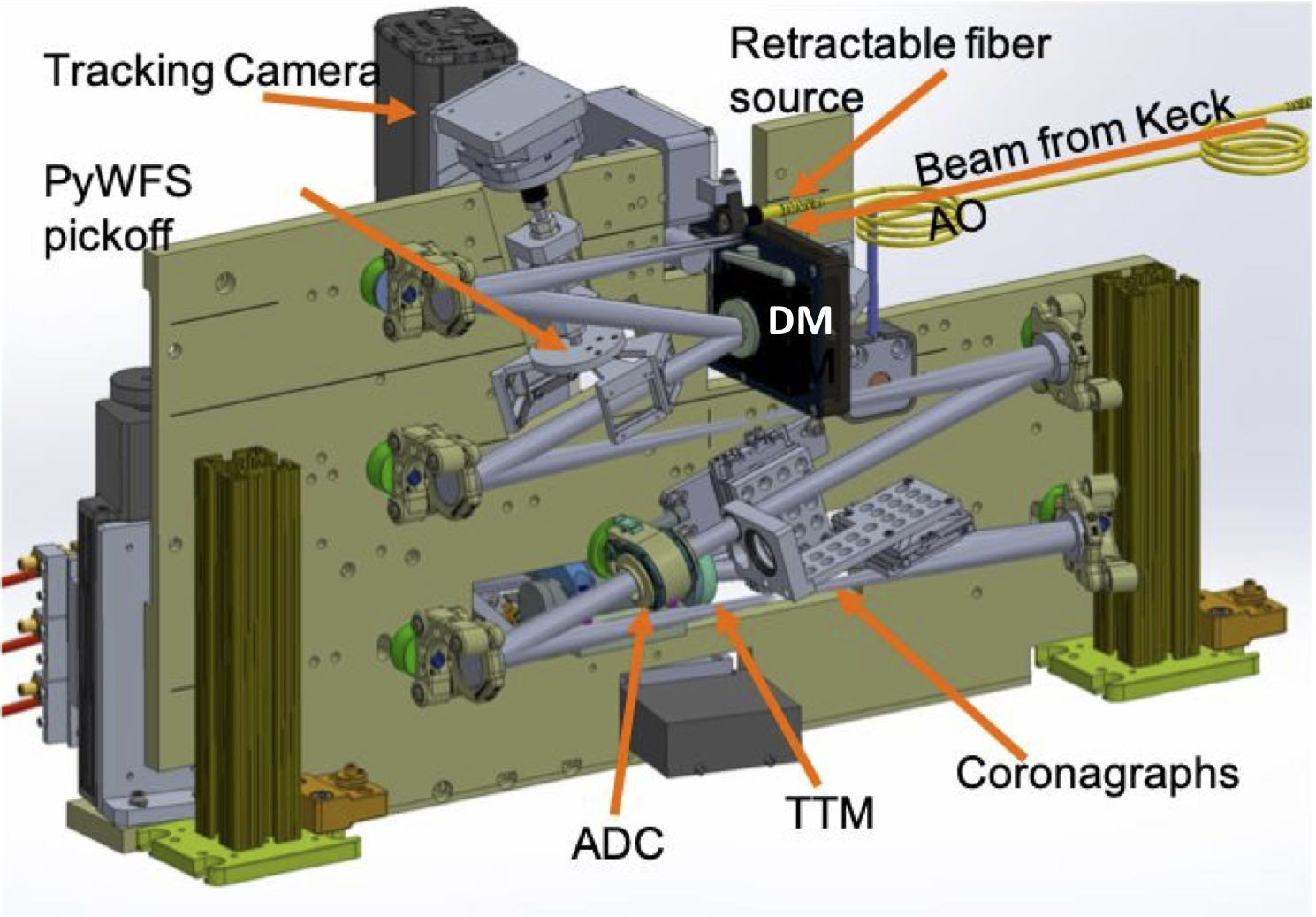} &
    \includegraphics[height=5.6cm]{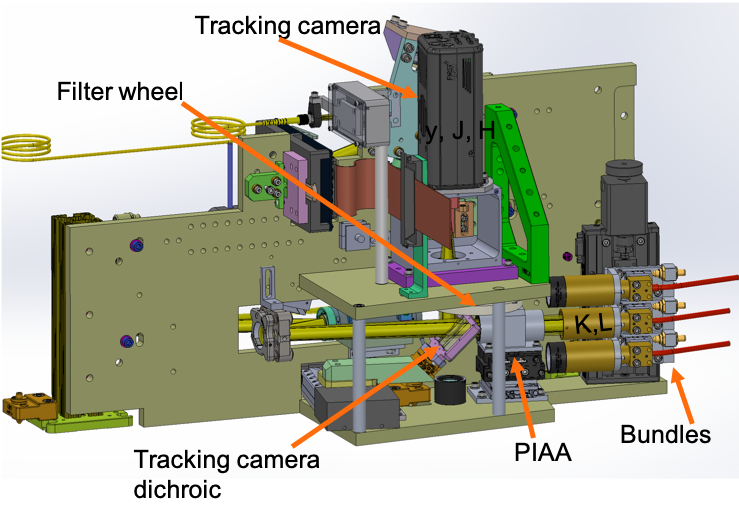}
    \end{tabular}
    \end{center}
    \caption[example] 
    { 
    \label{fig:kpic_cad} 
    CAD model of the KPIC phase II FIU plate. 
    The optical path is shown in gray.
    \textbf{Left:} The KPIC FIU accepts a beam from Keck AO or a retractable fiber source for calibrations.
    After hitting the 1K DM in the first pupil, the beam is split at the PyWFS pickoff dichroic.
    The coronagraphs submodule is in the second pupil, and the ADC is shortly after.
    The beam is transferred to the other side of the plate at the tip-tilt mirror (TTM).
    \textbf{Right:} After the TTM, the beam passes through the tracking camera dichroic.
    In the reflected arm, the light passes through a filter wheel before reaching the C-RED 2 tracking camera.
    In the transmitted arm, the beam passes through the PIAA submodule before reaching the fiber bundle port optimized for the current observation mode.
    }
\end{figure}

While the KPIC phase II upgrades will greatly improve the efficiency of the instrument, KPIC phase I is currently doing science on sky.
To ensure minimal disruptions to phase I operations, 
a testbed for KPIC phase II is being constructed.
Figure \ref{fig:testbed_cad} shows a CAD drawing of the testbed. 
It is analogous to the optical layout of the actual KPIC instrument that will be shipped to Keck, which is shown as a CAD model in Figure \ref{fig:kpic_cad} for comparison to the testbed.
Aside from being on flat plates, the only major differences between the testbed and the final KPIC optical layout stem from the decision to move the PyWFS pickoff location downstream of the ADC and coronagraphs submodules.
The decision to move this pickoff removes the need for additional optics in the testbed's PyWFS plate and simplifies the layout slightly.
The testbed will streamline integration of phase II into Keck by allowing for each of the new submodules to be fit tested in the KPIC optical layout, to be characterized within the system, and to have control software developed.

The KPIC phase II testbed will also incorporate a server, the Adaptive Optics Simulation Engine (AOSE), which mirrors the processes of the KPIC Real Time Controller (RTC) (Jensen-Clem et al. 2019, these proceedings).
The inclusion of AOSE will allow for the control scripts for each of the submodules to be developed in an environment identical to that in which they will operate at Keck.
AOSE currently controls the DM submodule, the tracking camera (a C-RED 2), and the PyWFS camera (Sensors Unlimited SU320HX-1.7RT).
Once the testbed is fully assembled, the predictive control simulations currently being developed on the AOSE machine will be tested with the KPIC system (Jensen-Clem 2019, these proceedings).

\begin{figure} [ht]
    \begin{center}
    \begin{tabular}{c} 
    \includegraphics[height=7cm]{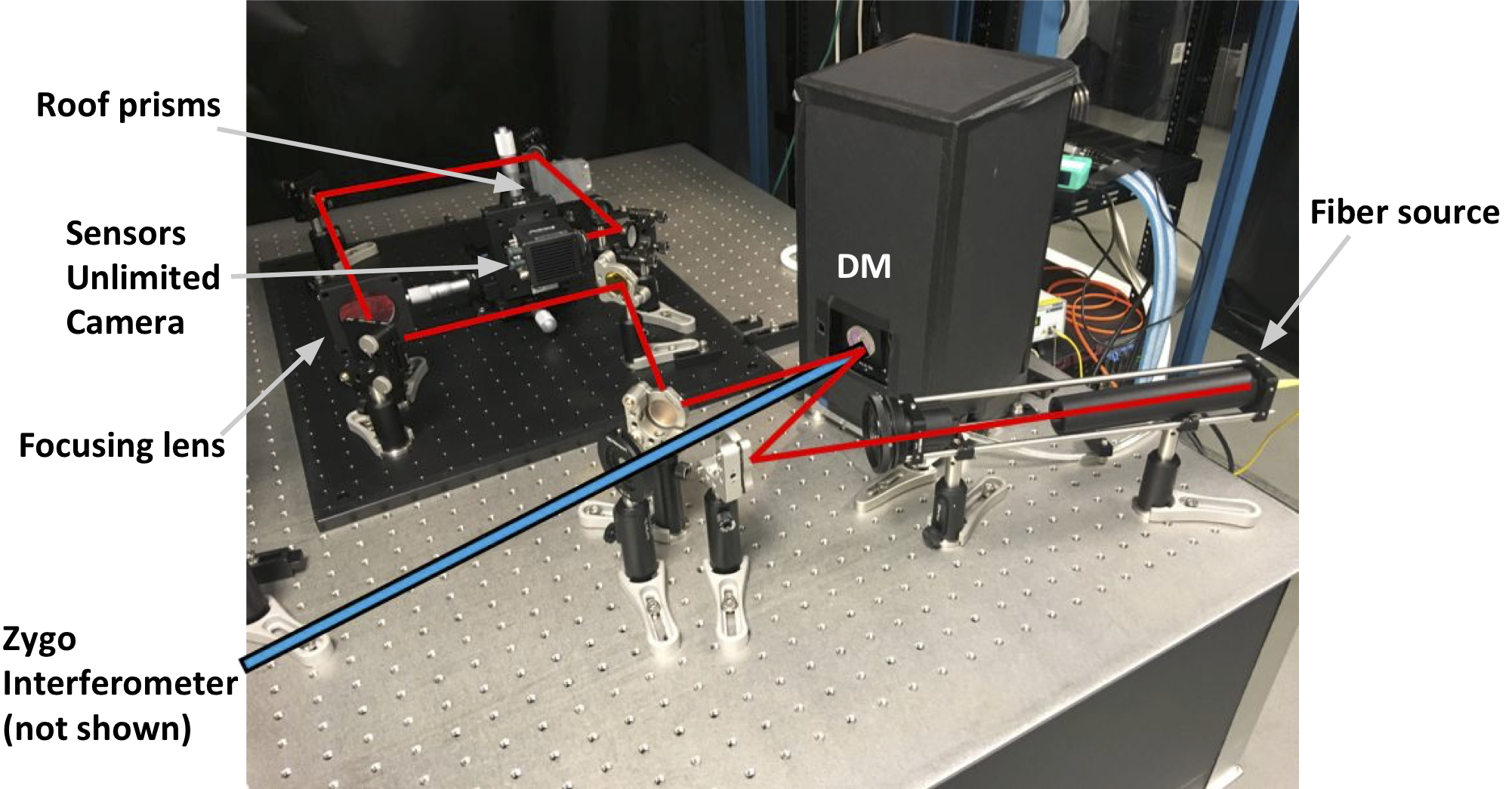}
    \end{tabular}
    \end{center}
    \caption[example] 
    { 
    \label{fig:testbed_img} 
    Picture of the KPIC phase II testbed as of August 2019.
    The PyWFS plate is fully aligned and its enclosure (not pictured) is built.
    The 1K DM is aligned with both the PyWFS plate and a Zygo Interferometer (not shown) for characterization.
    The Zygo beam (blue) is illuminating the DM, and an IR laser (red) is propagating through the PyWFS plate.
    }
\end{figure}

The construction of the testbed is currently underway.
The PyWFS plate is fully constructed as of August 2019.
As shown in Figure \ref{fig:testbed_img}, the PyWFS plate is also aligned with the KPIC DM, which is described in more detail in Section \ref{dm}.
The FIU plate has not yet been constructed, as many components have yet to arrive. 
However, some submodules are already undergoing characterization outside of the testbed before integration. 
More information on the status of the KPIC DM, coronagraphs, and PIAA lenses can be found in Sections \ref{dm_update}, \ref{corona_update}, and \ref{piaa_update}, respectively.

\section{The Deformable Mirror}
\label{dm}

\subsection{KPIC DM Design}
Phase II of KPIC includes the addition of a 1K continuous-surface Boston Micromachines (DM) into the path of the FIU. 
The new DM will have a 952-actuator surface, which will result in an improvement in Strehl compared to the 349 actuator Keck AO DM that is controlled in KPIC phase I.
The Strehl improvement may theoretically be between a factor of 1.4-1.13, depending on the band. 
By improving Strehl, the DM will improve raw contrast by decreasing the amount of contaminating starlight that is coupled into the SMF, since the amount of residual starlight is proportional to $\frac{1-S}{N^2}$, where $S$ is the Strehl ratio and $N$ is the total number of actuators.
The Strehl improvement also improves the coupling of the planet light into the SMF, therefore increasing throughput.
Thus, the Strehl improvement offered by the addition of the DM is expected to decrease the required exposure time to achieve a given signal-to-noise ratio (SNR), which can be found with the formula
\begin{equation}
    \label{eq:itime}
    t_{int} \propto \frac{\eta_s}{\eta_p^2}
\end{equation}
where $\eta_s$ is the coupling efficiency of the starlight and $\eta_p$ is the coupling efficiency of the planet light.
Figure \ref{fig:itime} shows the net effect of the DM and the submodules described in Sections \ref{coronagraphs}, \ref{adc}, and \ref{piaa} on the exposure time required to reach a given SNR relative to KPIC phase I, as calculated by this formula.

The window of the DM must be made with a material that is highly transmissive in the planned bands of operation for KPIC: \textit{y-M}.
The optical specifications have been achieved by careful selection of a custom AR-coated CaF$_2$ window, which has been mounted to the DM permanently. 
The window is not hermetically sealed 
and there is therefore a risk of long-term performance degradation of the DM if it operates regularly in a humid environment.
To mitigate this risk, the Keck Adaptive Optics bench will incorporate a low-pressure dry air line with a diffuser similar to that used in the  High-Resolution Spectroscopy for Segmented Telescopes (HCST-R) testbed at Caltech's Exoplanet Technology Lab (Llop Sayson et al. 2019, these proceedings). 
The dry air line is not expected to produce a significant amount of turbulence in the Keck Adaptive Optics Bench, since the diffuser is not near the beam and the air flow is very low.
Contrasts of 1x10$^{-8}$ have been demonstrated in HCST with a similar setup in place \cite{LlopSayson2019}.
There will also be a humidity sensor installed near the location of the DM in KPIC, and a DM electronics interlock will shut down the DM if the humidity in the bench spikes.

\subsection{KPIC DM Status}
\label{dm_update}
\begin{figure} [ht]
    \begin{center}
    \begin{tabular}{c} 
    \includegraphics[height=7cm]{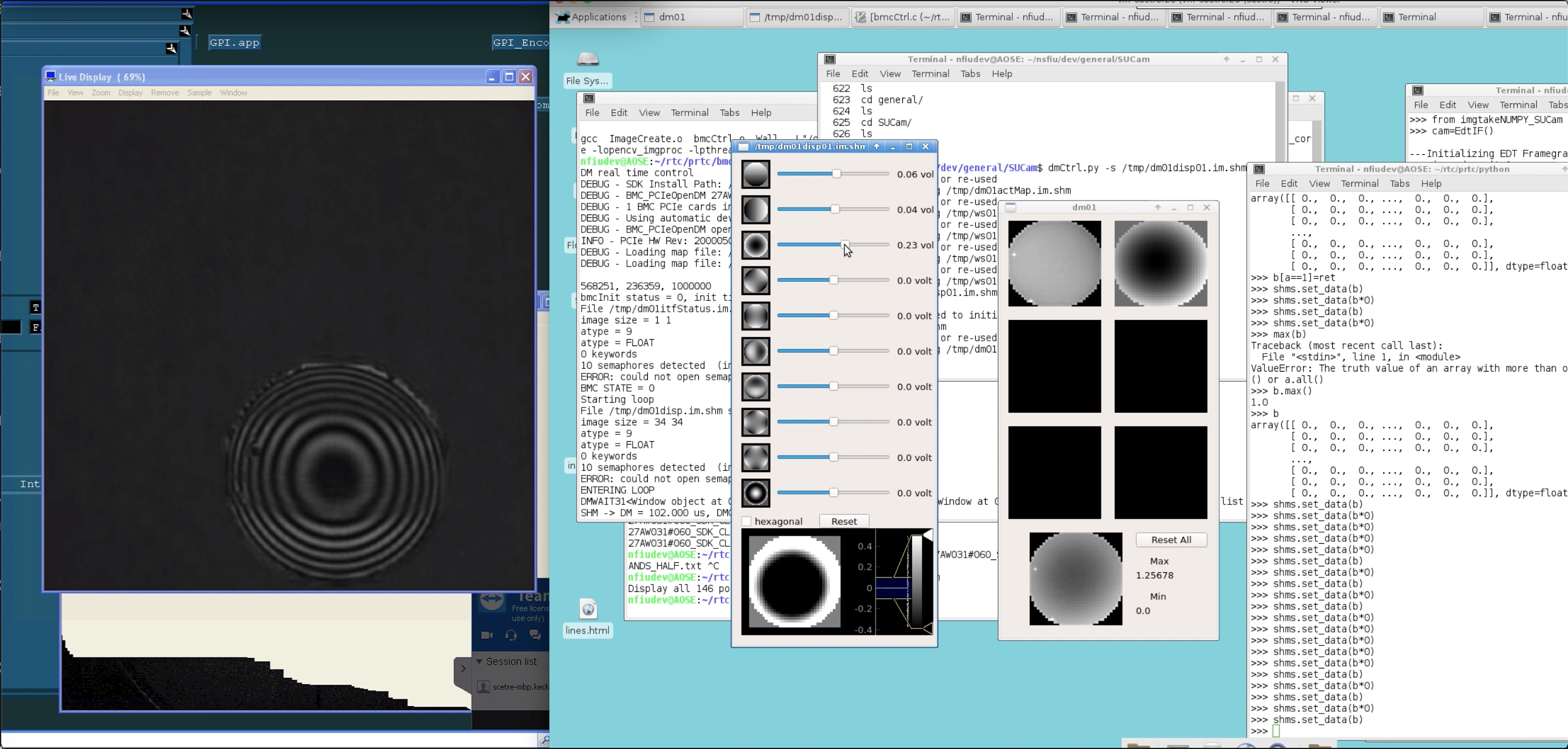}
    \end{tabular}
    \end{center}
    \caption[example] 
    { 
    \label{fig:dm_interface} 
    \textbf{Left:} 
    KPIC DM surface with defocus applied, as measured by the Zygo Interferometer.
    Fringes indicate a change in the optical pathlength.
    \textbf{Right:}
    The user interface for the KPIC DM, shown applying the Zernike mode that is being measured by the Zygo to the left.
    }
\end{figure}

As shown in Figure \ref{fig:testbed_img}, the KPIC DM is currently aligned with both the PyWFS plate of the KPIC testbed and a Zygo Interferometer beam.
The KPIC DM is controlled by a user interface shown in Figure \ref{fig:dm_interface}.
The development of this interface, which is similar to that used for the Keck AO DM, will simplify integration into Keck operations.
Also shown in Figure \ref{fig:dm_interface} is a screenshot of the Zygo Interferometer's output for the Zernike mode that the user is applying.
This user interface and the Zygo outputs are being used to characterize the KPIC DM within the system architecture in which it will be used.
Simultaneously, the PyWFS response to the applied Zernike modes can be calibrated.

\section{Coronagraphs}
\label{coronagraphs}

\subsection{Coronagraph Design}
KPIC phase II supports three science modes: direct spectroscopy in \textit{K-}band, direct spectroscopy in \textit{L-}band, and a Vortex Fiber Nulling (VFN) demonstration mode in \textit{K-}band (Echeverri et al 2019, these proceedings).
For targets limited by diffracted light from bright host stars, KPIC uses a custom pupil-plane microdot apodizer to produce a rotationally symmetric point spread function with attenuated wings.
The addition of this optic decreases the coupling efficiency of starlight into the fiber.
The VFN mode uses a charge 2 \textit{K-}band vortex coronagraph.
Observations that are limited by thermal background will be supported with PIAA lenses (see Section \ref{piaa}) in lieu of a coronagraph.

\begin{figure} [ht]
    \begin{center}
    \begin{tabular}{c} 
    \includegraphics[height=8cm]{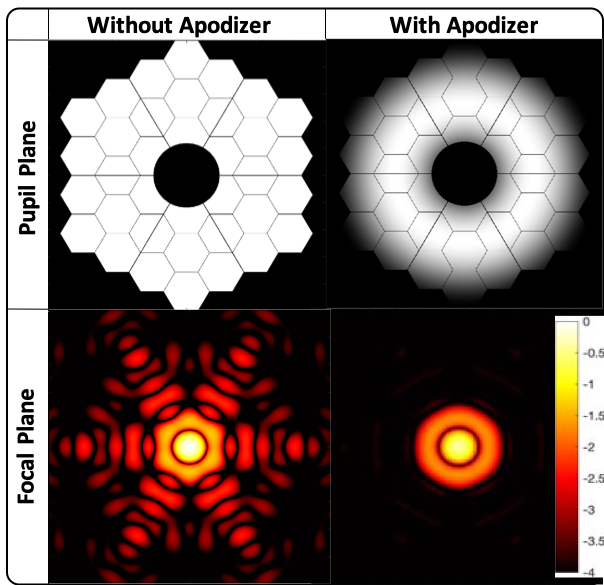}
    \end{tabular}
    \end{center}
    \caption[example] 
    { 
    \label{fig:apodizer_keck} 
    The second KPIC pupil and the corresponding point spread functions in the focal plane with and without the affect of the custom apodizer.
    }
\end{figure}

The apodizer mask is a custom chrome microdot pattern overlaid on a CaF$_2$ substrate.
The microdot pattern is made up of 25 $\mu$m x 25 $\mu$m chrome squares.
This microdot size was selected based upon results from Zhang et al. (2018)\cite{Zhang2018} that showed a linear relationship between local transmission and fill factor for a chrome-on-glass mask as long as the square dots are roughly $10\times$ the wavelength. 
The design of the mask minimizes the relative integration time expected in \textit{K-}band.
The performance was optimized for planet sources within 3-15 $\lambda/D$ from the star.
The radial apodization function was modeled as a 6th order polynomial.
The simulated effect of the final microdot pattern on the Keck PSF is shown in Figure \ref{fig:apodizer_keck}, along with the unapodized and apodized Keck pupil.

The charge 2 vortex coronagraph installed as a part of KPIC phase II will be used in an on-sky demonstration of VFN as a part of its third phase.
See Echeverri et al. 2019 (these proceedings) for more information on VFN in KPIC.

\subsection{Coronagraph Status}
\label{corona_update}
\begin{figure} [ht]
    \begin{center}
    \begin{tabular}{c} 
    \includegraphics[height=10cm]{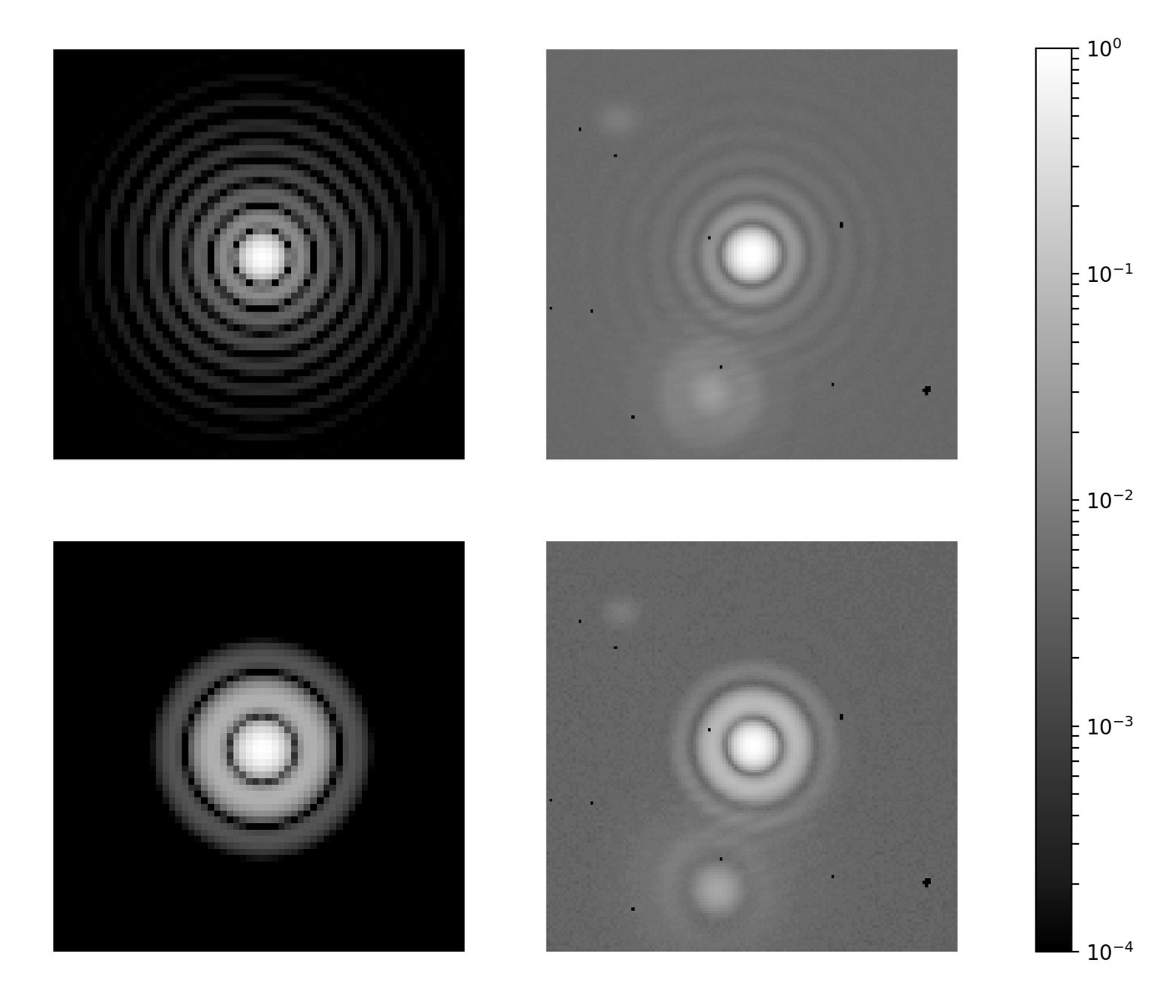}
    \end{tabular}
    \end{center}
    \caption[example] 
    { 
    \label{fig:apodizer_lab} 
    Simulated (left) and measured (right) \textit{K-}band PSFs with (bottom) and without (top) the apodizer mask with a circular unobstructed pupil.
    The lab data is background subtracted and normalized for easier comparison to the simulated PSFs.
    Note that the PSF measured with the apodizer has the expected number of rings.
    Two ghosts appear in the images collected in the lab.
    }
\end{figure}

The chrome microdot pattern for the apodizer mask has been successfully applied to a CaF$_2$ substrate.
The throughput and resulting PSF of the apodizer mask is currently being characterized outside of the KPIC testbed with a 2 micron laser that is collimated, sent through the optic, and then imaged by a NIR camera.
To verify that the apodizer is behaving as expected, the anticipated PSF with an unobstructed circular pupil is simulated in Matlab for comparison with measured results. 
Figure \ref{fig:apodizer_lab} shows the theoretical PSF and the PSF measured in the lab for the apodizer.
Note that the measured PSFs are similar in form to the theoretical PSF, and that suppression of the outer rings with the addition of the apodizer is observed.
Throughput tests are underway to verify that the apodizer is behaving as expected.

See Echeverri et al. 2019 (these proceedings) for information on the characterization of the KPIC charge 2 vortex in the lab.

\section{Atmospheric Dispersion Compensator}
\label{adc}

\subsection{ADC Design}
The atmosphere begins to refract the light of astronomical objects starting at the space/atmosphere boundary.
Light continues to refract as it propagates through the atmosphere, with the amount of refraction dependent on local atmospheric conditions (pressure, temperature, humidity, etc). 
Under a parallel-plane atmosphere assumption, the final amplitude of atmospheric refraction only depends on the index of refraction of air right at the primary mirror. 
This refraction is wavelength-dependent and results in significant elongation of a polychromatic PSF at large zenith angles.

Atmospheric dispersion causes two problems for KPIC.
The first issue is interband dispersion.
KPIC tracks in \textit{H-}band and does science in \textit{J-}, \textit{K-}, and \textit{L-}bands.
Atmospheric dispersion between bands greater than the size of the PSF in the tracking band therefore necessitates an offset in the tracking algorithm.
The size of this offset between the tracking and science band depends on the current atmospheric conditions (temperature, pressure, relative humidity) and the zenith angle of the target.
A small error in the tracking algorithm or offset calculation could result in the core of the science band being missed, resulting in a severe throughput penalty. 
Such blind offsetting has not yet been successfully demonstrated on sky, and incorporating an ADC submodule would mitigate much of the risks associated with interband dispersion, since a blind offset would not be required.

The second issue is intraband dispersion.
Significant elongation of the PSF in the science band will result in a decrease in the coupling efficiency of the light at the edges of the band.
This elongation could therefore result in a decreased SNR for molecular lines of interest, such as the CO band head at the edge of \textit{K-}band.
The addition of an ADC that corrects for intraband dispersion will therefore ensure that coupling efficiency is flat across the science band.

The tolerance for atmospheric dispersion is low for KPIC because of the need to couple the light into a SMF. 
To mitigate the affects of inter- and intra-band dispersion for KPIC, the ADC must correct atmospheric dispersion within and between the science and tracking bands to within 10\% of the diffraction-limited size of the PSF.
Therefore, KPIC's ADC must reduce the residual atmospheric dispersion to a few milliarcseconds across many bands (\textit{J-L}).


Careful selection of an atmospheric model is an essential aspect of ADC design. Since KPIC's ADC must correct for dispersion to within a few  milliarcseconds, the ADC is designed using Mathar's 2007\cite{Mathar2007} atmospheric model, which has been cross-checked at those scales with N-band on-sky observations\cite{Skemer2009} and another physical model of atmospheric diffraction (Guyon in prep.).
The model in Mathar (2007) accounts for water vapor and relative humidity, which seem to be ignored in these bands by the model currently used in Zemax, which is similar to the atmospheric model presented in Roe (2002)\cite{Roe2002}. 
Design of KPIC's ADC was therefore done outside of Zemax so that the Mathar (2007) model could be utilized.

\begin{figure} [ht]
    \begin{center}
    \begin{tabular}{c} 
    \includegraphics[height=5cm]{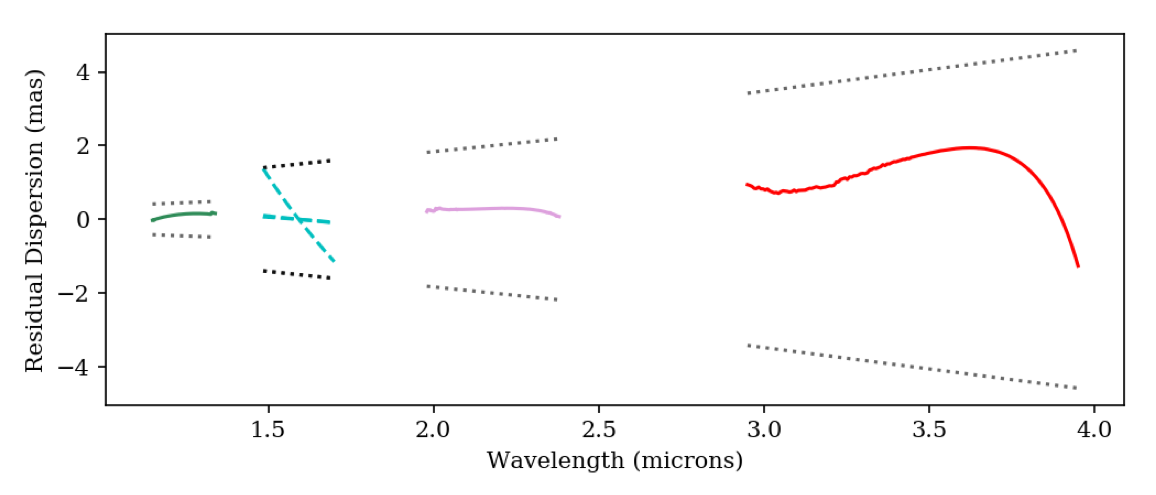}
    \end{tabular}
    \end{center}
    \caption[example] 
    { 
    \label{fig:adc_residual} 
    Performance of the ADC in \textit{J-}, \textit{H-}, \textit{K-}, and \textit{L-}band shown as residual dispersion relative to the center of \textit{H-}band after correction for a relative humidity of 20\%.
    All are shown for a target with a zenith angle of 60 degrees except \textit{J-}band, since the VFN \textit{J-}band science case is only defined up to a zenith angle of 30 degrees.
    The steeper blue line in \textit{H-}band corresponds to the residual tracking band dispersion in the \textit{L-}band science configuration, and the flatter \textit{H-}band line shows the residual tracking band dispersion when in the \textit{J-} and \textit{K-} band configurations.
    The black dotted lines at the edges of each band show the requirement of $\frac{1}{10}$\textsuperscript{th} the size of the diffraction-limited PSF at that wavelength.
    }
\end{figure}

The ADC was optimized in Python using an optical model that was verified with Zemax.
The optimization routine minimized four parameters for each of the science bands: the peak-to-valley (PTV) dispersion in the science band, the PTV dispersion in the tracking band, the offset between the tracking and science bands, and the deviation of a 1.55 $\mu$m ray through the ADC prisms.
The ADC consists of two sets of three cemented wedged prisms made of crystalline BaF$_2$, CaF$_2$, and ZnSe. 
The free parameters in the ADC optimization were the wedge angles of each of these materials and the clocking angle for each science band mode given the zenith angle and current atmospheric conditions.
The final ADC design meets the science requirements. 
Table \ref{table:adc} shows the residual dispersion at the maximum zenith angle for each of the KPIC phase II observing modes for the final ADC design.
Figure \ref{fig:adc_residual} shows this information graphically.

\begin{table}[h!]
\centering
\begin{tabular}{ |p{2cm}||p{2cm}|p{2cm}|p{2cm}|p{2cm}|p{2cm}|p{2cm}|  }
 \hline
 Mode & Zenith angle (degrees) & Science wavelength ($\mu$m) & Diffraction limit (mas) & Science band PTV dispersion (mas) & Offset to tracking band (mas) & Tracking band PTV dispersion (mas)\\
 \hline
 \textit{K-}band DS & 60 & 1.98-2.38 & 46.9 & 0.23 & 0.28 & 0.16 \\
 \textit{L-}band DS & 60 & 2.95-3.95 & 75.3 & 3.20 & 1.26 & 2.52 \\
 \textit{J-}band VFN & 30 & 1.15-1.34 & 27.1 & 0.22 & 0.14 & 0.21 \\
 \textit{K-}band VFN & 30 & 1.98-2.38 & 46.9 & 0.07 & 0.12 & 0.04 \\
 \hline
\end{tabular}
 \caption{ADC performance at 20\% relative humidity in KPIC phase II direct spectroscopy (DS) and VFN observing modes.
 Reported PTV dispersion is after correction.
 }
 \label{table:adc}
\end{table}

The KPIC ADC will be non-common to the PyWFS.
This choice is more convenient optomechanically and also prevents losses and ghosting from the ADC prisms in the pyramid's path.
If the ADC were common between the FIU and PyWFS, the ADC prisms could introduce complications, such as issues with the registration of the DM to the SAPHIRA and static aberrations in the WFS arm that would be noncommon to NIRC2 in imaging mode. 
This decision is supported by simulations of the effect of atmospheric dispersion on PyWFS performance which show no loss in the \textit{K-}band Strehl ratio when atmospheric dispersion is considered or when it is neglected \cite{Plantet2018}.

\subsection{ADC Status}
\label{adc_update}
The ADC has not been received as of August 2019.
Once the various ADC parts are received, the ADC will be characterized in the lab before being placed into the KPIC testbed.
Optical throughput will be measured within the testbed currently being used to characterize the coronagraphs, as described in Section \ref{corona_update}.
The net deliberate wedge angle and the non-common path abberations (NCPAs) as a function of clocking angle will be measured with the Zygo interferometer.

\section{Phase Induced Amplitude Apodization Lenses}
\label{piaa}

\subsection{PIAA Design}
The PIAA lenses are designed to reshape the centrally obscured flat-topped beam provided by the telescope to a quasi-Gaussian beam.
This quasi-Gaussian beam more closely matches the fundamental mode of the SMF at the location of the planet, therefore improving coupling of the planet light.
These optics also reshape the PSF of the on-axis stellar light and reduce the amount of background stellar light that couples to the fiber.
The addition of these optics therefore decreases the required amount of time to achieve a given SNR.

\begin{figure} [ht]
    \begin{center}
    \begin{tabular}{c} 
    \includegraphics[height=6cm]{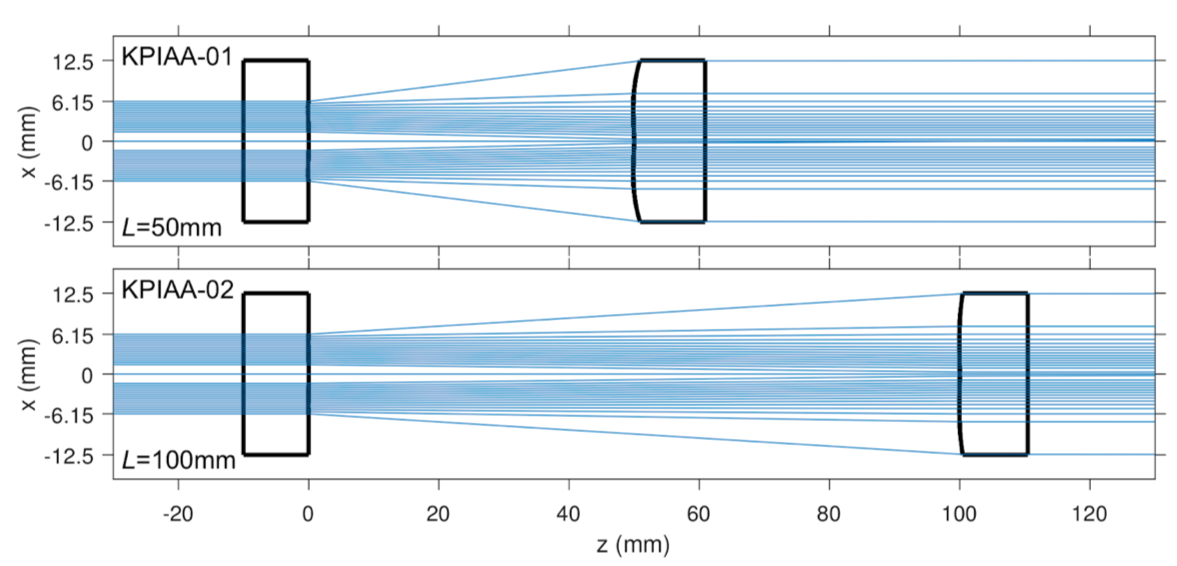}
    \end{tabular}
    \end{center}
    \caption[example] 
    { 
    \label{fig:piaa_ray} 
    Ray tracing diagrams for two possible designs for the KPIC PIAA lenses with different interlens distances.
    }
\end{figure}

\begin{figure} [ht]
    \begin{center}
    \begin{tabular}{c} 
    \includegraphics[height=7cm]{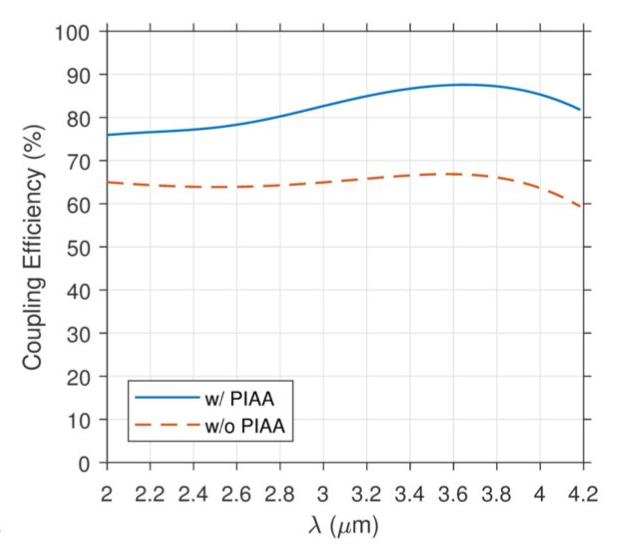}
    \end{tabular}
    \end{center}
    \caption[example] 
    { 
    \label{fig:piaa_coupling} 
    Theoretical coupling efficiency as a function of wavelength with and without the addition of the PIAA lenses.
    Note that the coupling efficiency is optimized for \textit{L-}band but that the PIAA still offers some improvement in \textit{K-}band. 
    The coupling efficiency is high across a large range of wavelength.
    }
\end{figure}

The two paired PIAA lenses each have one polished flat surface and one surface with a custom aspheric sag profile designed to reshape and recollimate the beam.
These two lenses are separated by an interlens distance, L.
Two possible PIAA lens pairs have been designed for KPIC.
One pair has a nominal interlens distance of 50~mm, and the other has with an interlens distance of 100~mm.
The lens pair with the smaller interlens distance would be preferable for its compact nature, but the pair with the larger interlens distance is less sensitive to alignment errors and beam size.
Figure \ref{fig:piaa_ray} shows the two proposed PIAA lens pairs in a ray tracing diagram.
Both designs offer improved coupling efficiency in \textit{L-} and \textit{K-}band, as shown in Figure \ref{fig:piaa_coupling}.
The PIAA lenses will predominantly be used when observations are limited by thermal background.
The PIAA submodule is optimized for \textit{L-}band, but also may be used for \textit{K-}band observations.

\begin{figure} [ht]
    \begin{center}
    \begin{tabular}{c} 
    \includegraphics[height=5cm]{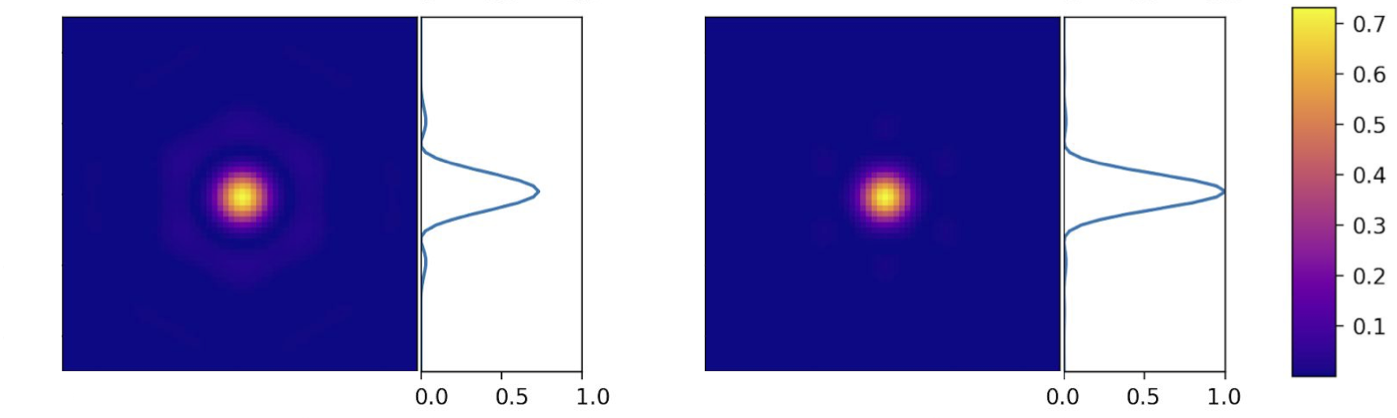}
    \end{tabular}
    \end{center}
    \caption[example] 
    { 
    \label{fig:piaa_keck} 
    Simulated PSF for the Keck pupil without (left) and with (right) PIAA lenses in the beam.
    To the side of each frame is a vertical cut through the core of each PSF.
    Note that the reshaped Keck PSF is closer in form to a symmetric 2D Gaussian, which will couple into the SMF more efficiently. 
    }
\end{figure} 


\subsection{PIAA Status}
\label{piaa_update}
The throughput and resultant PSFs of the PIAA lens pairs are being characterized in the 2 micron testbed described in Section \ref{corona_update}.
Again, the expected PSF for an unobstructed circular pupil can be simulated in Matlab so that the lab PSF can be used to verify the performance of the submodule without the Keck pupil.
Figure \ref{fig:piaa_keck} shows the theoretical PSF produced by the PIAA lenses in \textit{L-}band with the Keck pupil.


\section{Conclusions}
\label{conclusions}

\begin{figure} [ht]
    \begin{center}
    \begin{tabular}{c} 
        \includegraphics[height=6cm]{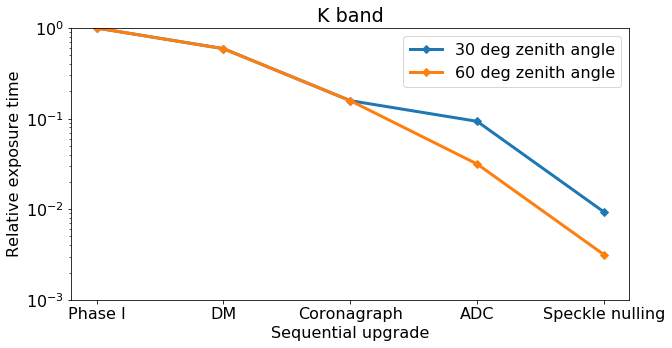} \\
        \includegraphics[height=6cm]{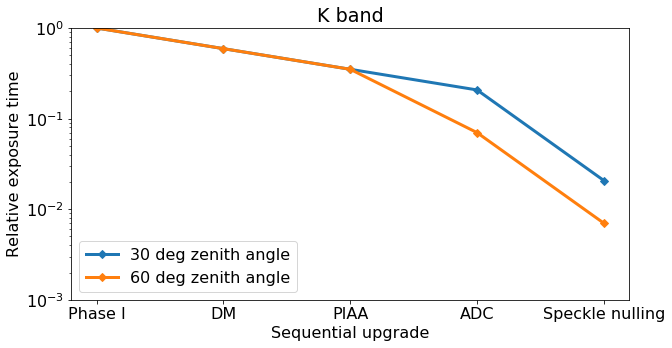} \\
        \includegraphics[height=6cm]{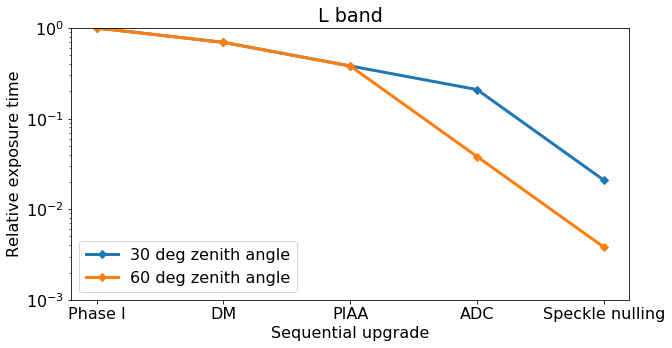}
    \end{tabular}
    \end{center}
    \caption 
    { 
    \label{fig:itime} 
    Incremental reduction in required exposure time necessary to achieve a given SNR on a target. 
    \textbf{Top:} \textit{K-}band spectroscopy mode with apodizer in the beam.
    \textbf{Middle:} \textit{K-}band spectroscopy mode with PIAA lenses in the beam.
    Although they are not optimized for \textit{K-}band, they still offer performance improvements.
    \textbf{Bottom:} \textit{L-}band spectroscopy mode with PIAA lenses in the beam.
    }
\end{figure} 

The addition of the four submodules detailed in this proceedings will greatly reduce the required exposure time to achieve a given SNR for targets observed with KPIC.
By improving Strehl, the addition of a 1K DM will both improve coupling of the planet light into the SMF and decrease the amount of residual starlight coupled into the SMF.
The DM may also allow for the implementation of techniques such as speckle nulling and electric field conjugation on sky, further improving raw contrast \cite{LlopSayson2019}.
For \textit{K-}band spectroscopy, the apodizer coronagraph decreases the amount of residual stellar light that is coupled into the SMF, having the net effect of increasing the SNR even with throughput losses for the planet. 
At high zenith angles, the ADC will decrease intraband dispersion, increasing the throughput of the planet light.
Finally, in both the \textit{K-} and \textit{L-}band spectroscopy observing modes, the PIAA lenses reshape the Keck pupil so that the planet PSF couples more efficiently into the SMF.
Figure \ref{fig:itime} summarizes the affect of each submodule on expected exposure time compared to KPIC phase I for L and \textit{K-}band spectroscopy, as calculated using Equation \ref{eq:itime}.
The net improvement offered by the addition of the submodules detailed in this proceedings is a reduction in the exposure time required to reach a given SNR of up to a factor of 333, depending on the observing mode, atmospheric conditions, and the target's zenith angle.


\acknowledgments      
The authors would like to acknowledge the financial support of the Heising-Simons foundation.
We thank Dr. Rebecca Jensen-Clem for loaning AOSE for use within the KPIC phase II testbed. 
Part of this work was carried out at the Jet Propulsion Laboratory, California Institute of Technology, under contract with the National Aeronautics and Space Administration (NASA).


\end{document}